# Flow Field Characteristics of a Supersonic Jet Influenced by Downstream Microjet Fluidic Injection

**Hamideh Pourhashem [a], Sunil Kumar [b], and Iraj M. Kalkhoran [a]**

[a] New York University, Tandon School of Engineering, Mechanical and Aerospace Engineering Department, 6 MetroTech Center, Brooklyn, NY 11201
[b] New York University Abu Dhabi, Division of Engineering, PO Box 129188, Saadiyat Island, Abu Dhabi, UAE

**ABSTRACT**

Mixing characteristics of a supersonic jet influenced by a Downstream Microjet Fluidic Injection (DMFI) system are numerically investigated. The DMFI system is built on the observation of a previous experimental study that utilized transverse fluidic injection from four equally spaced injection ports placed on an injection tube at a distance downstream of a 1.5 Mach number nozzle. The measurements from these previous experiments demonstrated thickening and mixing enhancement of the jet shear layer as a result of fluidic injection. The current numerical study examines the underlying physics of the flow field, as well as the effectiveness of the DMFI system at smaller mass flow rate ratios compared to those utilized in the previous experiments. Results indicate good agreement with the trend observed by the experimental study, and considerable improvement in enhancement of the jet mixing is observed. The observed mixing enhancement is attributed to the presence of the microjet tube and fluidic injection and the consequent generation of streamwise vortices, as well as the natural separating bow shock due to the transverse flow injection. DMFI system is shown to enhance the early mixing, resulting in attenuation of the downstream turbulence production. Furthermore, the DMFI system is demonstrated to be an effective method of mixing enhancement for supersonic jets with a potential for reducing the jet noise radiation.

## 1. Introduction

Enhancement of high speed jet mixing has various applications in the operational environment of jet engines such as promoting the supersonic combustion efficiency, and reducing the noise radiation from high-speed jets. Various mixing augmentation techniques developed in the past can be divided in two categories of passive and active control methods [1]. Passive methods generally involve a geometrical modification of the flow path such as the use of vortex generators in the form of tabs, cross-wire, or chevrons at the exit of the nozzle [2, 3]. In contrast, active flow control methods aim to dynamically alter the flow characteristics through a momentum or energy deposition into the flow. Examples of the active control methods are fluidic injection, and piezoelectric and plasma actuators [4-7].

Various mixing enhancement methods, which have been efficient for subsonic flows, demonstrated limited success for supersonic flows due to the low mixing rate and the compressibility effects in such flows [8]. Among the methods examined in the past, microjet fluidic injection is shown to be an effective method of mixing enhancement for both subsonic and supersonic flows. The effectiveness depends on the injector configuration and operating conditions including microjet diameter, injection angle and spacing, in addition to the momentum ratio and penetration depth of injectant into the cross flow [1, 9-15]. Alkislar *et al*. [16] performed microjet fluidic injection on a Mach 0.9 jet using 8 symmetrically arranged microjets at the exit of the nozzle with a penetration angle of 60° and a 0.4% mass flow rate ratio. They demonstrated that generation of streamwise vorticity due to the microjet injection is responsible for increasing the mixing, as well as the turbulence levels near the nozzle exit, which consequently results in reduction of downstream turbulence. Additionally, Arakeri *et al*. [17] investigated the effect of microjet injection on the flow field of a Mach 0.9 axisymmetric jet using 18 microjets at an angle of 45° and injection of 1.12% of the primary jet mass flux. They noticed an initial shear layer turbulence increase followed by a downstream turbulence reduction. Greska *et al.* [18] performed microjet fluidic injection for a Mach 1.38 jet, using microjet injection at an angel of 60° with mass flow rate ratios less than 2.2%. The results of their study showed a decrease in the shear layer turbulence levels. Another example of using microjets for mixing enhancement in supersonic jets is the work by Chauvet *et al*. [10], who performed a parametric study on mixing of supersonic round jets using radial injection near the nozzle exit. They achieved the maximum increase in the jet mixing efficiency through increasing the injection pressure ratio for the number of injectors being four or less. In a recent work by Hafsteinsson *et al.* [19], fluidic injection into a Mach 1.56 jet was investigated, utilizing 6 to 8 steady and flapping injectors at 90° angle with respect to the main jet flow. The results of their study indicated mixing enhancement with reduction of the mean axial velocity, as well as turbulent kinetic energy downstream of the jet potential core.

The present study examines the effect of microjet fluidic injection on supersonic jet mixing at a distance downstream of the nozzle exit. This is a different approach compared to the previous methods of fluidic injection, which were mostly limited to the injection within or at the exit of the nozzle. The fluidic injection method used in this study is a downstream 90° point injection through four equally spaced microjet ports at the extreme end of a tube placed inside the nozzle. The microjet tube can be traversed along the jet centerline downstream of the nozzle exit in order to effectively target the regions of the jet shear layer, which possess higher potential for mixing enhancement. This allows for targeting larger scales of turbulence, which govern the mixing process of the shear layer, in addition to an effective control of turbulence and the associated noise radiation from the convecting large turbulence scales. Large scale structures of the shear layer are sources of Mach wave emission for supersonic jets. Therefore, attenuating their propagation downstream can consequently reduce the Mach wave emitting region, and hence affect the jet noise radiation [20].

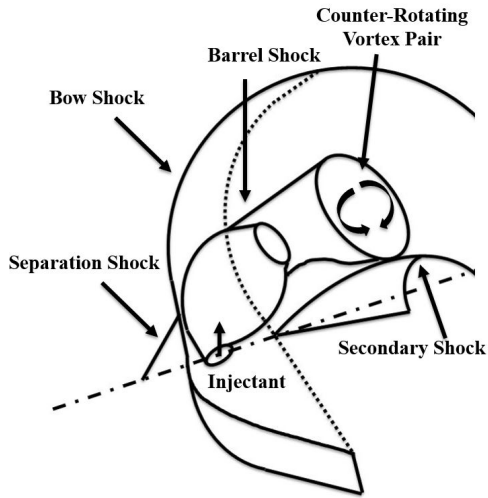

**Fig. 1: Schematic Sketch of fluidic injection into the supersonic flow**

In the present study, many of the well-known features of a sonic transverse injection into a supersonic cross flow are observed [21, 22]. Some of these features such as lambda, barrel, and bow shock, as well as counter rotating vortices are illustrated in Fig.1. Counter rotating vortices are known to enhance the jet mixing process by augmentation of the momentum exchange in the flow field [12, 16]. It is worth mentioning that the vortices generated by fluidic injection were observed to be more persistent in the flow field than those generated by the passive devices such as chevrons [9], and therefore have greater effect on modification of the mixing characteristics of the jet. Additionally, one of the notable features of the transverse flow injection into a supersonic cross flow is the formation of a bow shock upstream of the injection location. It is argued that the interaction of a sufficiently strong shock wave with the turbulent shear layer of the jet plume can increase the vorticity [23], which consequently influences the evolution and propagation of the vortical structures of the shear layer downstream. Previous studies of the transverse flow injection into a supersonic cross flow [24] have shown that the strength of the generated bow shock is influenced by the momentum ratio of the injecting jet into the supersonic cross flow, which per se depends on the dynamics of the flow at the injection location [25-29] . As a consequence, a higher jet momentum ratio gives rise to a stronger bow shock and potentially a more pronounced alteration of the shear layer mixing.

The objectives of the present numerical study are to investigate how Downstream Microjet Fluidic Injection (DMFI) system alters the stability characteristics of the shear layer through incorporating and varying the parameters, which could increase the possibility of mixing enhancement, such as variation of injection location ($L/D$) that allows for targeting the jet and the shear layer within the regions of high susceptibility to mixing alteration. In addition, through variation of mass flow rate ratio ($m_r$), the effect of injectant momentum ratio on mixing characteristics of the supersonic jet is examined. It should be noted that evaluation of the jet noise reduction capability of the proposed fluidic injection arrangements is not in the scope of the present article.

## 2. Problem Statement

Similar to the previous experimental study by Caeti and Kalkhoran [4], as well as the numerical studies by Pourhashem and Kalkhoran [25, 26], the supersonic jet flow used in the present numerical investigation is produced from a converging-diverging nozzle. The nozzle used in the present investigation has a design Mach number of 1.5 and an exit diameter (*D*) of 70 mm. The nozzle is operated at a pressure ratio (NPR) of 4, which corresponds to an under-expanded condition and an exit Reynolds number of approximately $2.5 \times 10^6$. For the DMFI equipped nozzle (Fig.2), the tube is placed along the nozzle centerline and extended from the nozzle inlet to a distance *L* downstream of the nozzle exit, with the tube to nozzle exit diameter ratio (*d/D*) of 0.1. Additionally, there are four equally spaced microjet ports, which are 90° apart and located near the extreme end of the tube, each with an exit diameter of 0.02*D*.

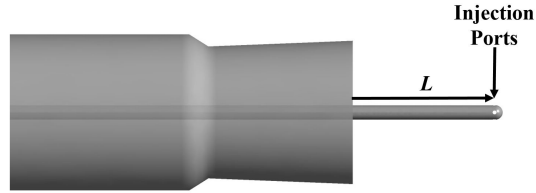

**Fig. 2: Geometric representation of DMFI system**

In the present study, air is used as both the injection and primary gas mediums. Also, similar to the experimental arrangements, the injection tube can be traversed along the nozzle centerline in order to examine the effect of injection location on the flow field characteristics. Furthermore, the flow rates through the nozzle and the injection tube can be independently varied. The results are compared for two different locations corresponding to *L/D* = 0.5 and *L/D* = 1, with ratio ($m_r$) of the total microjet mass flow rate to the main jet mass flow rate of 0.1% and 0.2%.

## 3. Computational Methodology

### 3.1. Governing Equations

In the present study a three-dimensional Delayed-Detached Eddy Simulation (DDES) [30-32] incorporating Realizable k-ε turbulence model [33] is implemented. In this methodology, the turbulence model is switched from an Unsteady Reynolds Averaged Navier-Stokes (URANS) model for the attached boundary layers near the wall regions to a Large Eddy Simulation (LES) [34] away from the walls. A finite volume CFD code is utilized, which solves the governing equations for the conservation of continuity, momentum, and energy [35] (all the variables in the following mathematical development have their usual meaning):

$$\frac{\partial}{\partial t}(\bar{\rho}) + \frac{\partial}{\partial x_i}(\bar{\rho}\tilde{u}_i) = 0 \tag{1}$$

$$\frac{\partial}{\partial t}(\bar{\rho}\tilde{u}_i) + \frac{\partial}{\partial x_j}(\bar{\rho}\tilde{u}_i\tilde{u}_j) = -\frac{\partial \bar{p}}{\partial x_i} + \frac{\partial}{\partial x_j}(\tilde{\tau}_{ij} + \tau_{ij}) \tag{2}$$

$$\frac{\partial}{\partial t}(\bar{\rho}\tilde{E}_0) + \frac{\partial}{\partial x_j}((\bar{\rho}\tilde{E}_0 + \bar{p})\tilde{u}_j) = -\frac{\partial}{\partial x_j}(q_{Lj} + q_{Tj}) + \frac{\partial}{\partial x_j}(\tilde{u}_i(\tilde{\tau}_{ij} + \tau_{ij})) \tag{3}$$

while the gas is considered to be thermodynamically perfect:

$$\bar{p} = \bar{\rho} R \tilde{T} \tag{4}$$

In the above equations $\bar{\rho}$ and $\bar{p}$ are the density and pressure, while $\tilde{u}_i$ and $\tilde{T}$ represent the velocity and temperature. Furthermore, the heat flux terms in the energy equation are given as:

$$q_{Lj} + q_{Tj} = -C_p \left(\frac{\mu}{P_r} + \frac{\mu_t}{P_{rt}}\right) \frac{\partial \tilde{T}}{\partial x_j} \tag{5}$$

The stress tensor term due to molecular viscosity, $\tilde{\tau}_{ij}$, and the subgrid-scale stress tensor formulation, $\tau_{ij}$, are defined as:

$$\tilde{\tau}_{ij} = \mu\left(\frac{\partial \tilde{u}_j}{\partial x_i} + \frac{\partial \tilde{u}_i}{\partial x_j}\right) - \frac{2}{3}\left(\mu \frac{\partial \tilde{u}_k}{\partial x_k}\right)\delta_{ij} \tag{6}$$

$$\tau_{ij} = \mu_t\left(\frac{\partial \tilde{u}_j}{\partial x_i} + \frac{\partial \tilde{u}_i}{\partial x_j}\right) - \frac{2}{3}\left(\bar{\rho} k + \mu_t \frac{\partial \tilde{u}_k}{\partial x_k}\right)\delta_{ij} \tag{7}$$

The transport equations of kinetic energy and the dissipation rate for the Realizable k-ε turbulence model are given as:

$$\frac{\partial}{\partial t}(\bar{\rho}k) + \frac{\partial}{\partial x_j}(\bar{\rho}k\tilde{u}_j) = \frac{\partial}{\partial x_j}\left[\left(\mu + \frac{\mu_t}{\sigma_k}\right)\frac{\partial k}{\partial x_j}\right] + G_k - \bar{\rho}\varepsilon - Y_M \tag{8}$$

$$\frac{\partial}{\partial t}(\bar{\rho}\varepsilon) + \frac{\partial}{\partial x_j}(\bar{\rho}\varepsilon\tilde{u}_j) = \frac{\partial}{\partial x_j}\left[\left(\mu + \frac{\mu_t}{\sigma_\varepsilon}\right)\frac{\partial \varepsilon}{\partial x_j}\right] - \bar{\rho} C_2 \frac{\varepsilon^2}{k + \sqrt{v\varepsilon}} \tag{9}$$

where $G_k$ presents the turbulent kinetic energy production as a result of mean velocity gradients, and $Y_M$ is the dilatation dissipation term.

$$G_k = \mu_t \tilde{S}^2 \tag{10}$$

$\tilde{S}$ and $\mu_t$ are the modulus of the strain rate tensor and the turbulent eddy viscosity, respectively. These terms are defined as:

$$\tilde{S} \equiv \sqrt{2\tilde{S}_{ij}\tilde{S}_{ij}}, \quad \mu_t = \bar{\rho} C_\mu \frac{k^2}{\varepsilon} \tag{11-12}$$

where

$$\tilde{S}_{ij} = \frac{1}{2}\left(\frac{\partial \tilde{u}_j}{\partial x_i} + \frac{\partial \tilde{u}_i}{\partial x_j}\right) \tag{13}$$

The constants of the model in Eqs. 8 and 9 are modified in order to improve the accuracy of the Realizable $k$-$\varepsilon$ model for prediction of turbulent jet flows, as proposed by Thies and Tam [36]:

$$C_2 = 2.02, \quad \sigma_k = 0.324, \quad \sigma_\varepsilon = 0.377, \quad Pr_t = 0.422 \tag{14}$$

A blending function is implemented in the DDES model in order to ensure that the solution remains Unsteady RANS (URANS) near the wall region, even if the grid spacing becomes smaller than the boundary layer thickness:

$$f_d = 1 - \tanh((20\, r_d)^3) \tag{15}$$

with

$$r_d = \frac{v_t + v}{\sqrt{U_{ij}U_{ij}} k^2\, l_{rk\varepsilon}^2} \tag{16}$$

where $v$ is the kinematic eddy viscosity, $U_{i,j}$ the velocity gradients, $k$ the Kármán constant, and $l_{DDES}$ is defined as:

$$l_{DDES} = l_{rk\varepsilon} - f_d \max(0,\ l_{rk\varepsilon} - C_{DDES}\Delta)\ \text{with}\ \Delta = \max(\Delta x, \Delta y, \Delta z),\ \text{and}\ C_{DDES} = 0.61 \tag{17}$$

In the above equation, $l_{rk\varepsilon} = k^{3/2}/\varepsilon$ is the RANS length scale. The DDES length scale, $l_{DDES}$, is substituted in the dissipation term of the $k$ equation as shown below:

$$\frac{\partial}{\partial t}(\bar{\rho}k) + \frac{\partial}{\partial x_j}(\bar{\rho}k\tilde{u}_j) = \frac{\partial}{\partial x_j}\left[\left(\mu + \frac{\mu_t}{\sigma_k}\right)\frac{\partial k}{\partial x_j}\right] + G_k - \bar{\rho}\frac{k^{3/2}}{l_{DDES}} - Y_M \tag{18}$$

Thus, the DDES formulation switches to a Realizable $k$-$\varepsilon$ turbulence model near the wall regions, while far from the wall it becomes similar to a LES model with a subgrid turbulent kinetic energy model. As such, in the separated regions, the $\bar{\rho}k^{3/2}/l_{DDES}$ term becomes larger than the one in the Realizable k-$\varepsilon$ turbulence model, and hence, results in reduction of the turbulent kinetic energy without introducing a dissipation term.

In the present study, the numerical formulation is switched from a second order upwind to central difference between RANS and DDES regions, and the convective fluxes of the momentum equation are evaluated by a Roe flux-difference splitting scheme. The gradients of the solution variables at cell centers are determined using a least squares cell-based gradient evaluation, and a second order fully implicit time scheme is used to march the solution in time. The non-dimensional simulation time step ($\Delta t c_\infty/D_j$) is 0.003, with the CFL number being less than 0.2. The flow simulation runs for the total duration of nearly 300 non-dimensional time ($T c_\infty/D_j$) to achieve a statistically converged solution.

### 3.2. Test Cases

Two different test cases are considered to compare the results of the present numerical methodology with those from the previous experimental studies. The first test case is conducted using the baseline nozzle (i.e., no tube) of the experimental study by Caeti and Kalkhoran [4], while maintaining the same operating conditions. Figure 3 compares the schlieren image of the baseline case, obtained from the aforementioned experimental study (upper half of the figure), to the instantaneous density gradient contours of the present investigation (lower half of the figure). Typical of supersonic nozzles operating at underexpanded condition, expansion waves are formed at the nozzle exit, followed by a series of compression and expansion waves downstream of the nozzle exit. Figure 3 indicates that the location of the expansion waves emanating from the nozzle lip, as well as the location of the shock fronts interacting with the shear layer are predicted well for $x/D < 3$. Additionally, the predicted growth rate of the shear layer is in agreement with the one observed in the schlieren image of the experimental study. For both cases, a considerable shear layer roll-up is seen in close proximity to $x/D = 3$. Despite the lack of clarity of the schlieren image for $x/D > 3$, which makes further comparisons difficult, good agreement between the results of the numerical and experimental investigations is observed.

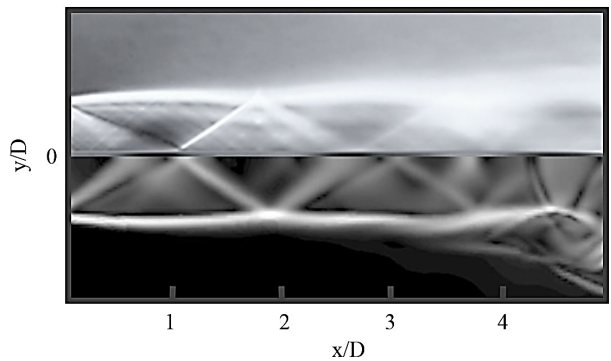

**Fig. 3: Comparison of the density gradient contours of the numerical investigation (lower half) to the schlieren image from the experimental study of Caeti and Kalkhoran [4] (upper half) for the baseline case**

To further investigate the accuracy of our numerical methodology, a DDES of an almost-perfectly expanded jet flow corresponding to the SMC015 nozzle is carried out (Fig.4). Details of the nozzle geometry can be found in the literature [37, 38]. The operating conditions of the nozzle correspond to an unheated jet with a jet Mach number of nearly 1.4. Figure 4 shows the instantaneous jet velocity profile (U/Uj) for SMC015 nozzle using the DDES methodolgy. For this validation case, a grid spacing corresponding to 0.01D is used in the shear layer region of the jet. It should be noted that a higher grid resolution near the nozzle lip can improve the results and capture the fine scale turbulent structures of the jet shear layer more accurately. As also outlined by Spalart [31], an inadequate grid resolution can result in a slow development of LES in the shear layer leading to non-resolved turbulence structures. As seen in Fig.5, the mean and root mean square velocity (RMS) profiles of the DDES method are compared with the experimental results of Bridges and Wernet [37], as well as the LES results of Mendez *et al.* [38] along the centerline and lipline of the nozzle. Comparison of the mean velocity results indicates an accurate prediction of the jet core length compared to the experiment. However, there is a minor under-prediction of shock cell strength, which is also observed in the LES results by Mendez *et al*. [38]. This inconsistency could be due to uncertainties associated with the pressure mismatch at the nozzle exit between the jet and the ambient, as well as the nozzle boundary layer of the computational study compared to the experiment [38]. Additionally, examination of the profiles of the RMS velocity shows some level of discrepancy in the results compared to the experiment as slightly higher turbulent fluctuation values are predicted by the DDES. Overall, comparison of the results demonstrates that the most fundamental features of the flow field including mean and turbulent characteristics are adequately captured by the DDES model.

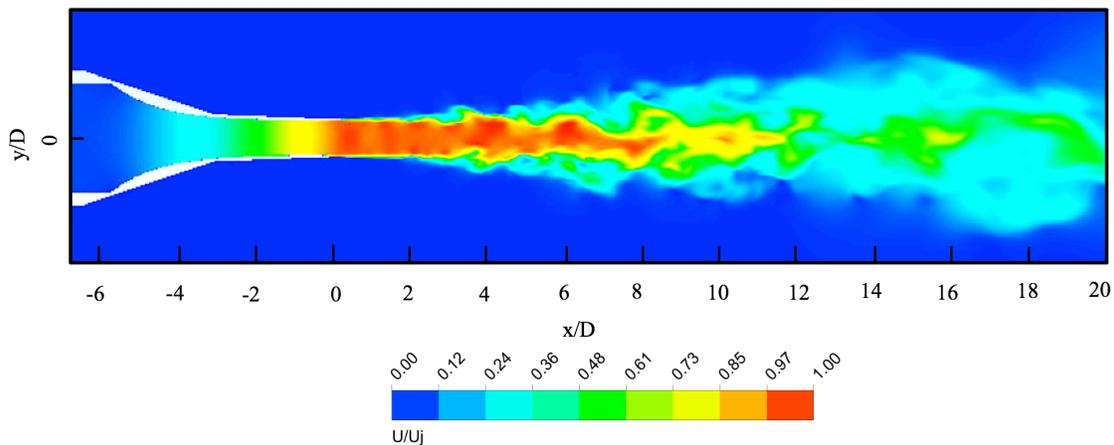

**Fig.4: Instantaneous jet velocity profile (U/Uj) for SMC015 nozzle using the DDES methodology (Uj corresponds to the fully expanded jet velocity)**

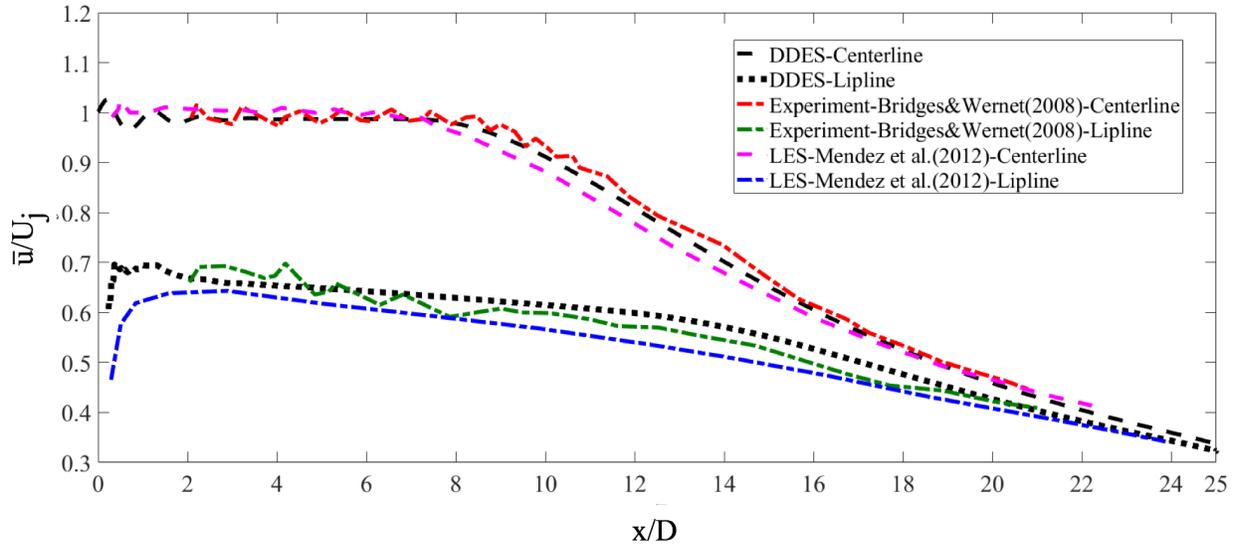

*(a) Mean velocity*

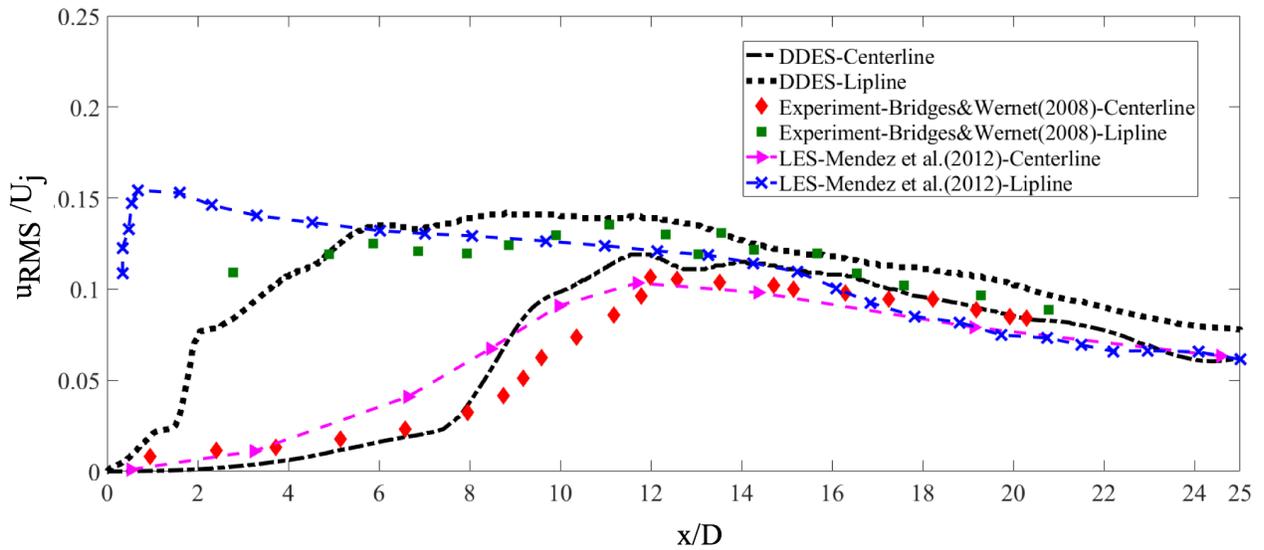

*(b) RMS Velocity*

**Fig.5: Comparsion of the mean and RMS velocity profiles of the DDES with the experimental results by Bridges & Wernet [37] and LES of Mendez et al. [38]**

### 3.3. Computational Domain

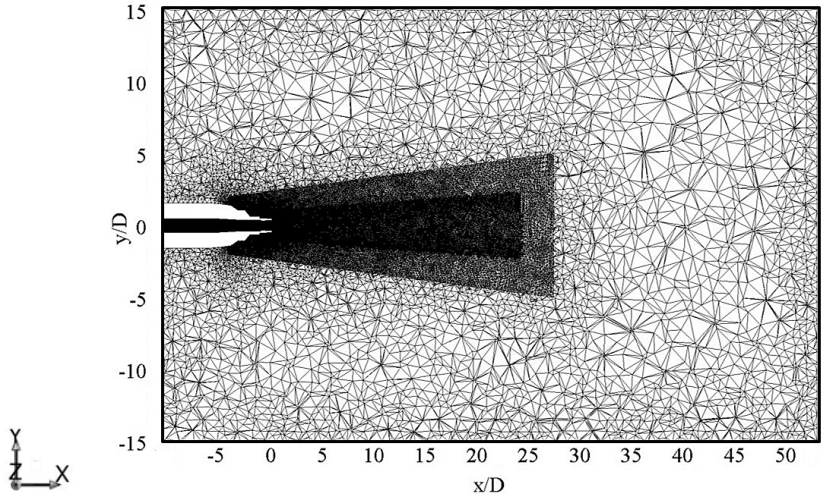

*(a) Computational domain in y-x plane at z/D=0*

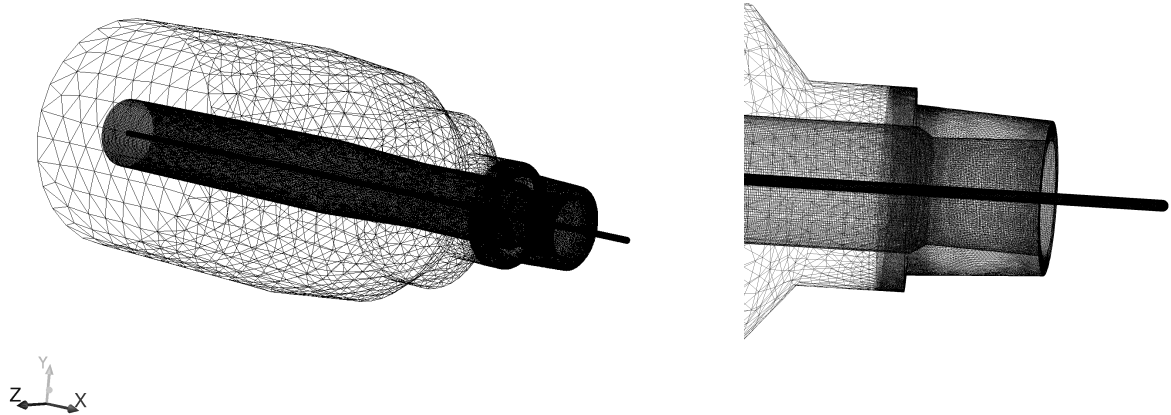

*b) DMFI equipped nozzle and the surrounding domain*   *(c) Close-up view of the DMFI equipped nozzle*

**Fig. 6: Computational domain**

Figure 6 illustrates the computational domain of the present study that includes nearly 53$M$ elements. The refined regions, which are shown in Fig. 6(a) include the nozzle interior and the region outside of the nozzle close to the jet core. The mesh refinement in the domain was limited to the aforementioned regions in order to maintain a reasonable computational demand, while achieving the desirable accuracy. Finer cells that are located at and upstream of the nozzle exit are nearly 0.009$D$ in size. The mesh size is increased downstream of the nozzle exit, reaching 0.026$D$ in the jet core region, which is then kept at this resolution for nearly 16D. The mesh size is further increased to 0.04D and is maintained at this size up to x/D = 25, which is then coarsened for x/D > 25. Figures 6(b) and 6(c) show the computational mesh for the DMFI equipped nozzle and the surrounding domain. As seen in Fig.6 (c), quadrilateral elements are used inside the nozzle, while a tetrahedral mesh type is applied for the regions outside of the nozzle. The reason for selecting the tetrahedral elements for the region outside of the nozzle is that this type of mesh is more practical for complicated geometries, such as in our case with a microjet tube.

A constant total pressure value is specified for the boundary condition at the nozzle inlet, as well as at the microjet injection ports, while the total temperature ratio is maintained equal to unity. The defined static pressure at the microjet injection inlets allows the flow to exit the injection ports at a sonic condition. Additionally, the ambient condition is applied at the far-field boundaries and a no-slip boundary condition is imposed on all solid surfaces in the domain, including the nozzle walls and the tube surface. The boundary layer on the inner walls of the nozzle and the microjet tube surface are modeled using the RANS approach with $y^+ \sim 1$ [26]. Although this approach does not resolve all details of the nozzle boundary layer, it nevertheless introduces adequate jet inflow condition to the domain without using artificial inflow forcing.

### 3.4. Grid Sensitivity Study

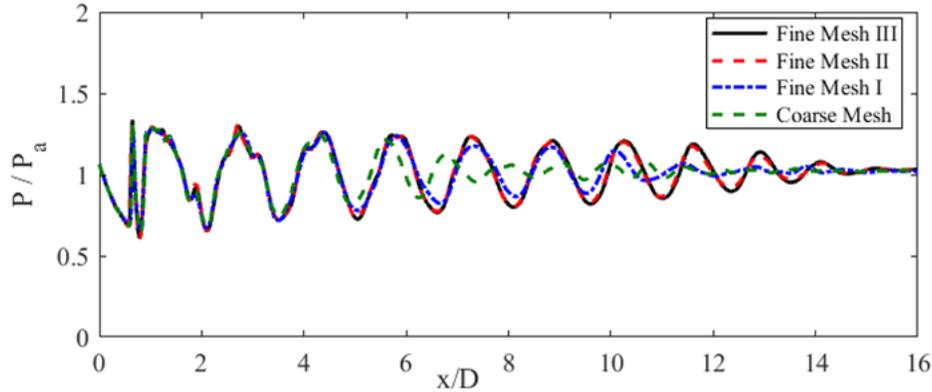

**Fig.7: Grid sensitivity study showing pressure profile of the baseline case along y/D=0 (Pa is the ambient pressure)**

In order to capture the fundamental features of the flow field based on the grid independency, a grid sensitivity study is carried out. The grid sensitivity corresponds to the jet flow of the nozzle, which was previously described in Section 2. As seen in Fig. 7, the static pressure distributions along the nozzle centerline are compared for four mesh resolutions. A coarse mesh is initially used, with nearly 27M grid cells and 0.05D spacing for the jet core region, as well as for the upstream and downstream regions of the nozzle exit. The mesh is refined, and the grid spacing is reduced to $0.04D$ (Fine mesh I). As seen in Fig.7, this refinement results in prediction of a longer jet core, as well as stronger shock cells. In an attempt to examine the precision of the results, the grid spacing is reduced to $0.026D$ in the jet core region, and to $0.01D$ in the regions extending to $1D$ and $4D$ downstream and upstream of the nozzle exit. This grid refinement increases the number of grid cells to nearly $48M$ (Fine mesh II), which in turn, results in a considerable increase in the jet core length. To evaluate the grid independency of the results, the mesh is further refined and the grid spacing is reduced to nearly $0.009D$ for the regions extending $1D$ and $4D$ downstream and upstream of the nozzle exit, respectively, with the total number of elements reaching approximately $53M$ cells (Fine mesh III). It is evident that the flow field predictions of fine mesh III are very similar to fine mesh II. The jet core length is predicted almost identical with minor increase in the strength of the shock cells downstream. Hence, with the result demonstrating a sufficient grid independency and considering the computational cost that is associated with any further grid refinements, the grid resolution corresponding to fine mesh III is selected and used for the present numerical investigation.

## 4. Results and Discussions

The characteristics of the transverse microjet fluidic injection are studied and discussed in this section, in addition to the examination of the effect of DMFI on the turbulent and mean characteristics of the flow field. The results are presented for 7 cases, including a baseline case (no microjet tube) and the cases for the DMFI equipped nozzle with two different microjet tube lengths, corresponding to $L/D = 0.5$ and $L/D = 1$, each with no injection, as well as injection with $m_r = 0.1\%$ and $0.2\%$.

## 4.1. Dominant Features of Transverse Fluidic Injection into a Supersonic Cross Flow

Figure 8 illustrates Schlieren images obtained from the experimental study by Caeti and Kalkhoran [4]. Those cases involving microjet tubes with both injection and without are presented in the upper halves of Fig. 8, while the

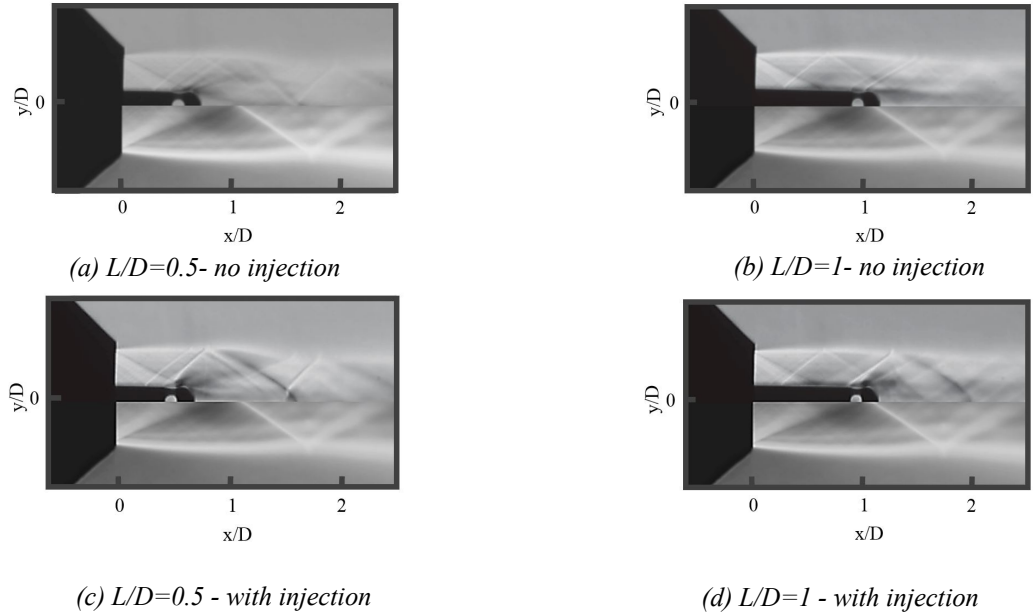

*(a) L/D=0.5- no injection*   *(b) L/D=1- no injection*

*(c) L/D=0.5 - with injection*   *(d) L/D=1 - with injection*

**Fig.8: Schlieren images of the experimental study (d/D=0.25) [4]. The lower halves of the figures represent the baseline case while the upper halves show cases with microjet tube**.

baseline case is shown in the lower halves. The baseline corresponds to the jet without the tube [4]. Different patterns of jet flow development and shock cell structure are observed for the DMFI equipped nozzles compared to the pattern for the baseline case. When the microjet tube is present, expansion waves emanating from the nozzle lip interact with the boundary layer of the tube surface and strike the shear layer in the form of shock waves, introducing additional shock structures into the jet flow [26]. In addition to the aforementioned phenomena, presence of the microjet tube leads to boundary layer separation and formation of a wake region immediately downstream of the tube end. As observed from Figs. 8(c) and 8(d), with activation of the fluidic injection, further shock structures including a small barrel shock as well as the bow shock are observed in the vicinity of the injection location. As discussed earlier, the generated bow shock interacts with the shear layer of the supersonic jet and affects the mixing characteristics of the shear layer. This trend, which is also observed in the present numerical study, is discussed later in this section.

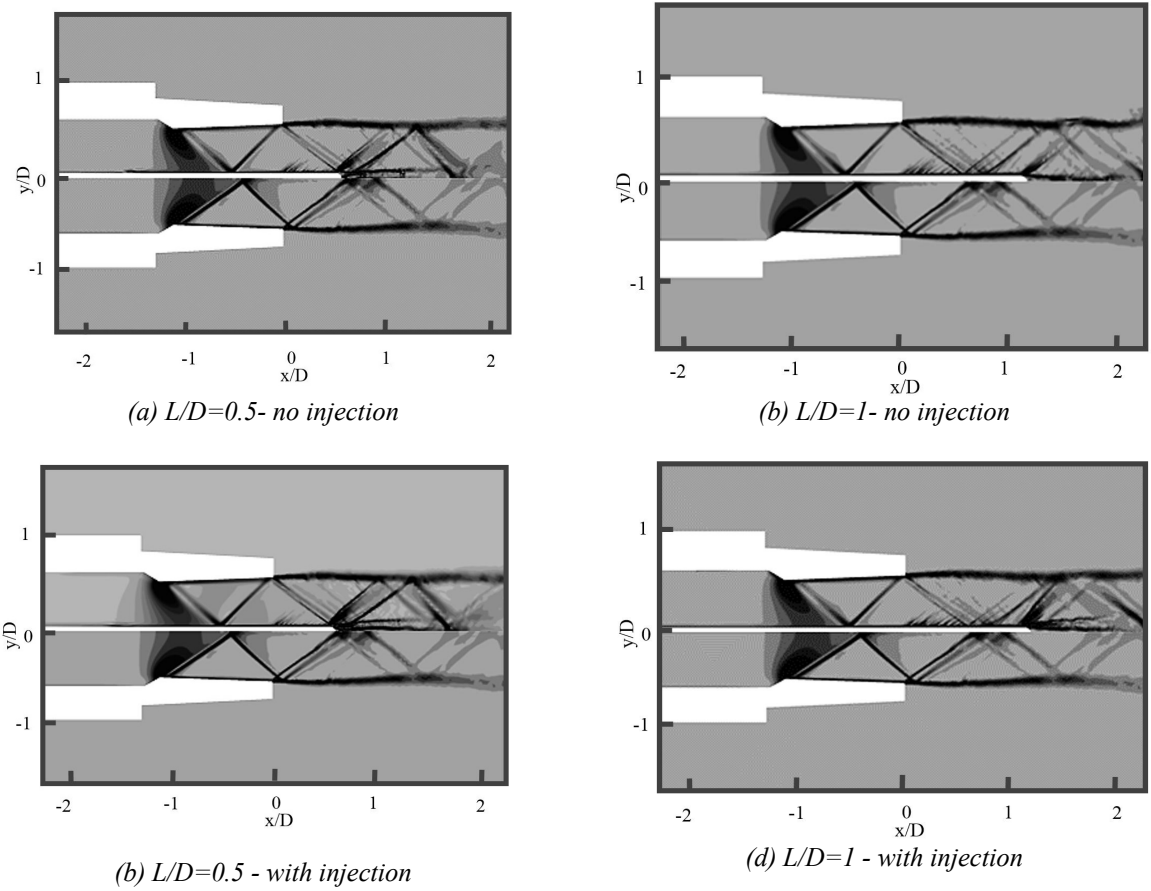

*(a) L/D=0.5- no injection*  *(b) L/D=1- no injection*

*(b) L/D=0.5 - with injection*  *(d) L/D=1 - with injection*

**Fig. 9: Comparison of the density gradient contours of the baseline case in the present numerical study to those involving the microjet tube (d/D=0.1) [26]. The lower halves of the figures represent the baseline case without presence of the tube, while the upper halves show cases with microjet tube.**

Contour plots of instantaneous density gradient obtained from the present numerical study are compared in Fig. 9 for two injection locations and two mass flow rate ratios. The bottom half of each figure is the computed baseline case that does not have the tube present. As seen in upper halves of Fig. 9, the flow patterns for the cases involving the microjet tube, with and without fluidic injection, are similar for the regions inside the nozzle and upstream of the injection point. The density gradient contours indicate formation of the expansion waves at the tube end, which reflect from the shear layer in the form of compression waves. Similar to the results of the experimental study, a bow shock is produced due to the transverse fluidic injection, which strikes the shear layer and is reflected in the form of expansion waves. Additionally, a wake region is formed downstream of the tube end, which is expected to influence the jet flow and shock cell structures through formation of wake vortices, which induce momentum exchange.

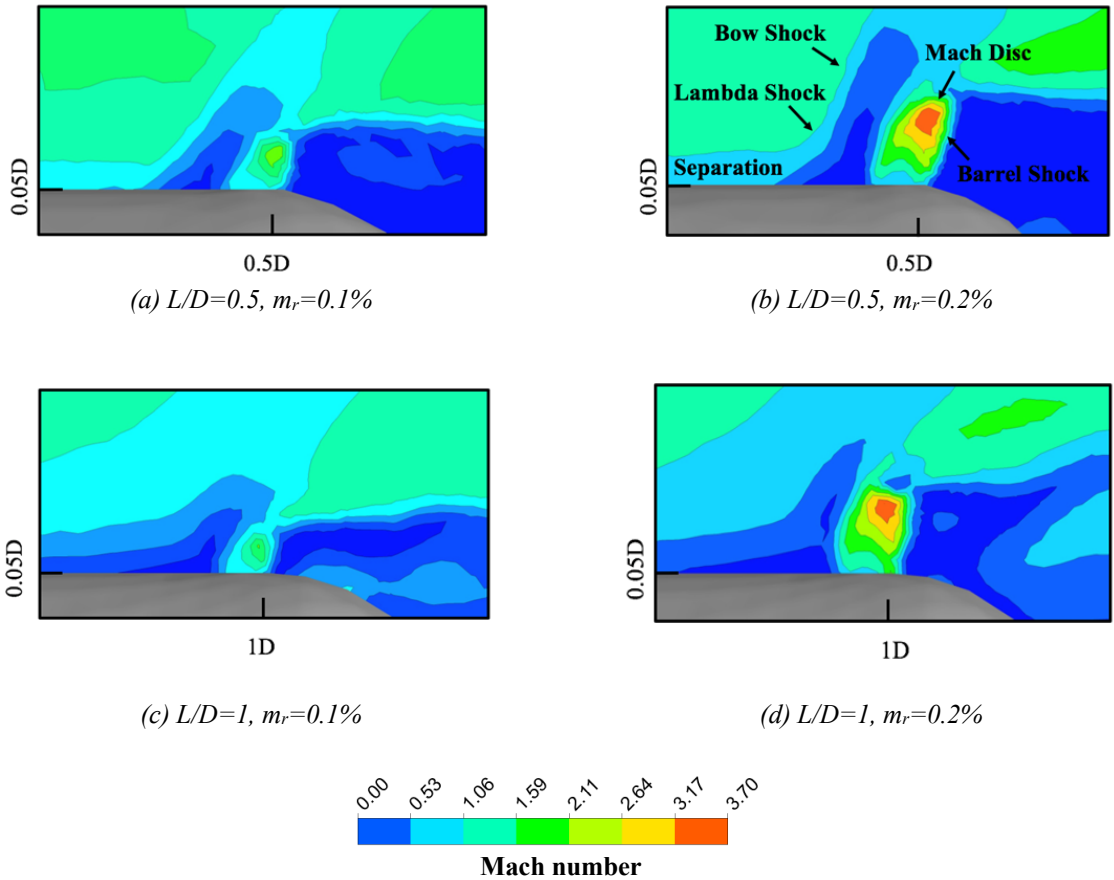

**Fig.10: Comparison of the underexpanded microjet injecting jet in x-y plane for L/D=0.5 and L/D=1**

To closely examine the character of the flow field in the vicinity of the transverse injection, the contour plots of Mach number in *x-y* plane are shown in Fig. 10. Figure 10 demonstrates the effect of injection location on the dynamics of the jet flow, where many of the well-known characteristics of the underexpanded transverse injection into the supersonic cross flow [21, 22, 27, 28] are visible. As seen in the Mach number contours of Fig. 10, the boundary layer separation over the tube just upstream of the injection is followed by formation of a lambda shock merging into the bow shock. Furthermore, formation of the barrel shock and expansion of the injected gas just below the Mach disk is also evident. As observed from Fig. 10, the injected gas flow at *L/D* = 0.5 expands to a slightly greater Mach number for both mass flow rate ratios compared to the injections at *L/D* = 1. Additionally, the size and inclination of the barrel shocks are not identical for the aforementioned cases, which is due to the distinct dynamics of the jet flow at the injection locations. As previously seen in Fig. 9, the injection port at *L/D* = 0.5 is in close proximity to the expansion waves that originated at the nozzle exit, whereas the one at *L/D* = 1 is within the region, where the compression waves are reflected from the shear layer. Additionally, comparing the structure of the flow field for two mass flow rate ratios demonstrates that increasing the momentum ratio of the injected gas influences the shape of the barrel shocks and increases the strength and steepness of the upstream bow shocks [24].

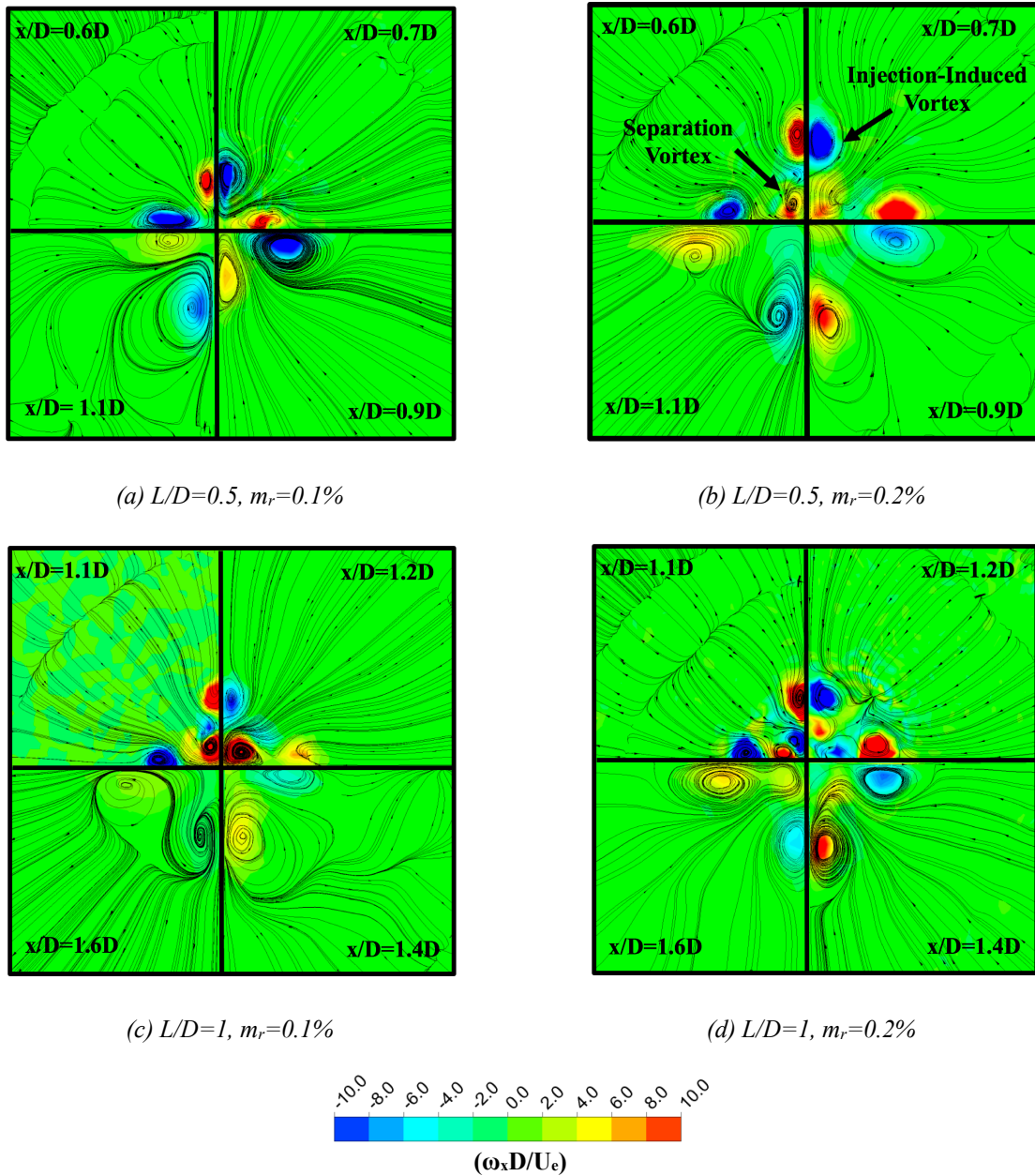

Fig. 11: Contours of Streamwise non-dimensional axial vorticity ($\omega_x D/U_e$) in y-z plane. The figure illustrates the evolution of the streamwise counter-rotating vortices, in a clockwise order, downstream of the injection points for L/D=0.5 and L/D=1.

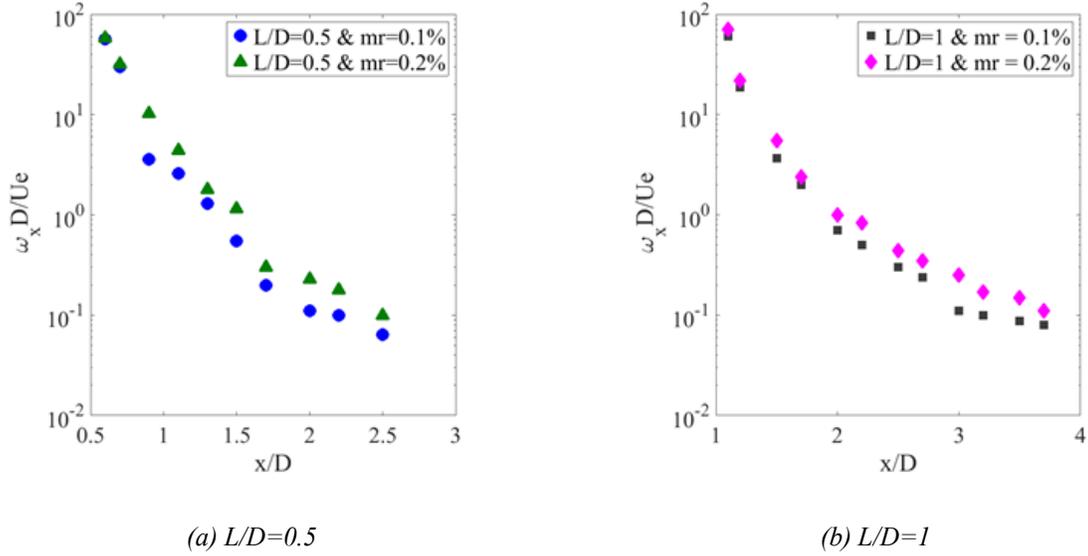

*(a) L/D=0.5*  *(b) L/D=1*

**Fig. 12: Development of streamwise non-dimensional axial vorticity ($\omega_x D/U_e$) due to transverse microjet fluidic injection ($U_e$ is the jet velocity at the nozzle exit).**

As noted in the previous section, as a result of transverse injection, counter-rotating vortices are formed downstream of the injection location, which are known to affect the mixing characteristics of the jet by promoting the momentum exchange in the flow field [1, 26]. Figure 11 shows the contour plots of streamwise vorticity ($\omega_x D/U_e$) in the *y-z* plane for injection at $L/D = 0.5$ and $L/D = 1$. Evolution of streamwise vortices for $L/D = 0.5$ are shown in Figs. 11(a) and 11(b) at four axial locations corresponding to $x/D = 0.6$, $x/D = 0.7$, $x/D = 0.9$ and $x/D = 1.1$. Similarly, Figs. 11(c) and 11(d) demonstrate the streamwise vortices due to the injection at $L/D = 1$, for $x/D = 1.1$, $x/D = 1.2$, $x/D=1.4$ and $x/D = 1.6$. As shown in Fig.11, counter-rotating vortices due to both injection as well as flow separation are visible in close proximity to the microjet tube end. Further downstream, the separation vortices dissipate all together, while the counter rotating vortices grow in size before diminution. As these vortices grow in size, they induce momentum exchange in the jet plume and their magnitude decays in the streamwise direction [26]. Figure 12 compares the magnitude of these generated streamwise vortices for both injection locations up to a distance downstream, where their magnitude becomes negligible compared to the initial values. It is evident from Fig.12 that the generated vortices associated with $m_r = 0.2\%$ retain higher magnitudes for a longer distance downstream of the injection points compared to those due to the injection with $m_r = 0.1\%$.

## 4.2. Turbulent Characteristics of the Flow Field

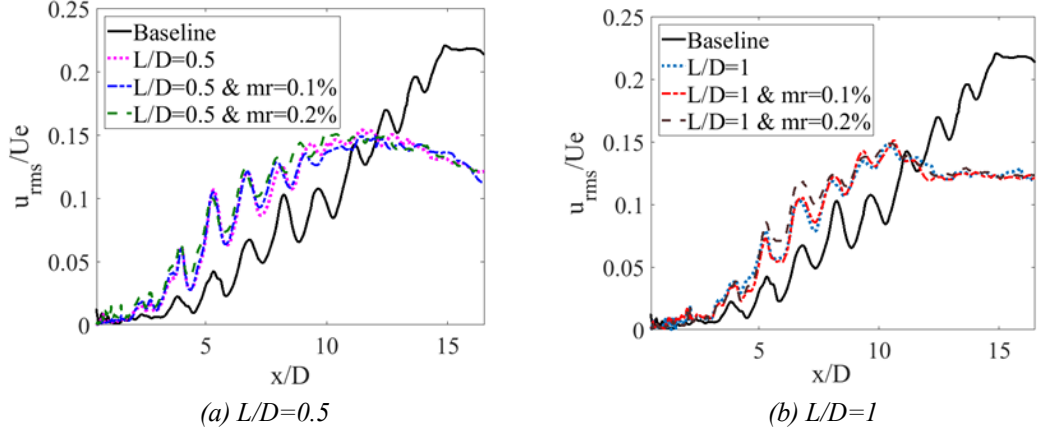

(a) L/D=0.5

(b) L/D=1

**Fig. 13: Profile of Turbulence Intensity close to the centerline at y/D=0.1**

To better understand the influence of microjet fluidic injection on the development of turbulent structures, profiles of the streamwise turbulence intensity are examined in Fig. 13. As seen in Figs. 13(a) and 13(b), the levels of turbulent intensity for all cases involving the microjet tube (with and without injection) are higher compared to the baseline case up to nearly $x/D = 11$. Beyond this location, turbulence intensities decrease to values lower than those for the baseline case. Furthermore, and as seen in Fig. 13, the peaks of turbulence intensity move upstream due to the presence of the microjet tube and fluidic injection, indicating shortening of the jet potential core length. For a global view of turbulence distribution in the flow field, the profiles of non-dimensional Turbulent Kinetic Energy (T.K.E) are presented in Figs. 14 and 15. These figures compare the T.K.E profiles for the baseline case and the cases involving microjet tube at four streamwise locations corresponding to $x/D = 4$, $x/D = 8$, $x/D = 12$, and $x/D = 16$. Turbulent kinetic energy is calculated from the components of turbulent velocity as:

$$\text{T.K.E} = \left[\frac{1}{2}\left(u_x'^2 + u_y'^2 + u_z'^2\right)\right]^{\frac{1}{2}} \tag{19}$$

As seen in Figs. 14 and 15, in addition to the increased levels of T.K.E, turbulence profiles are widened at both $x/D = 4$ and $x/D = 8$ for the cases with microjet tube compared to the baseline case. Further downstream at $x/D = 12$, T.K.E peaks remain nearly constant for the cases involving DMFI compared to those at $x/D = 8$, while the turbulence levels continue to increase for the baseline case. At $x/D = 16$, a decrease in the T.K.E levels are observed for all cases, with the baseline case demonstrating maximum values compared to other cases. Similar to the trends seen for the profiles of streamwise turbulence intensity, increased levels of turbulence are observed in the region near the nozzle exit for the cases with microjet tube compared to the baseline case. This implies that DMFI enhances the fine scale mixing near the nozzle exit, while attenuating the propagation of large coherent turbulence structures downstream [1]. As a result of enhanced fine mixing and upstream dissipation, downstream turbulence production is mitigated, which is also reflected in the profiles of T.K.E. [16].

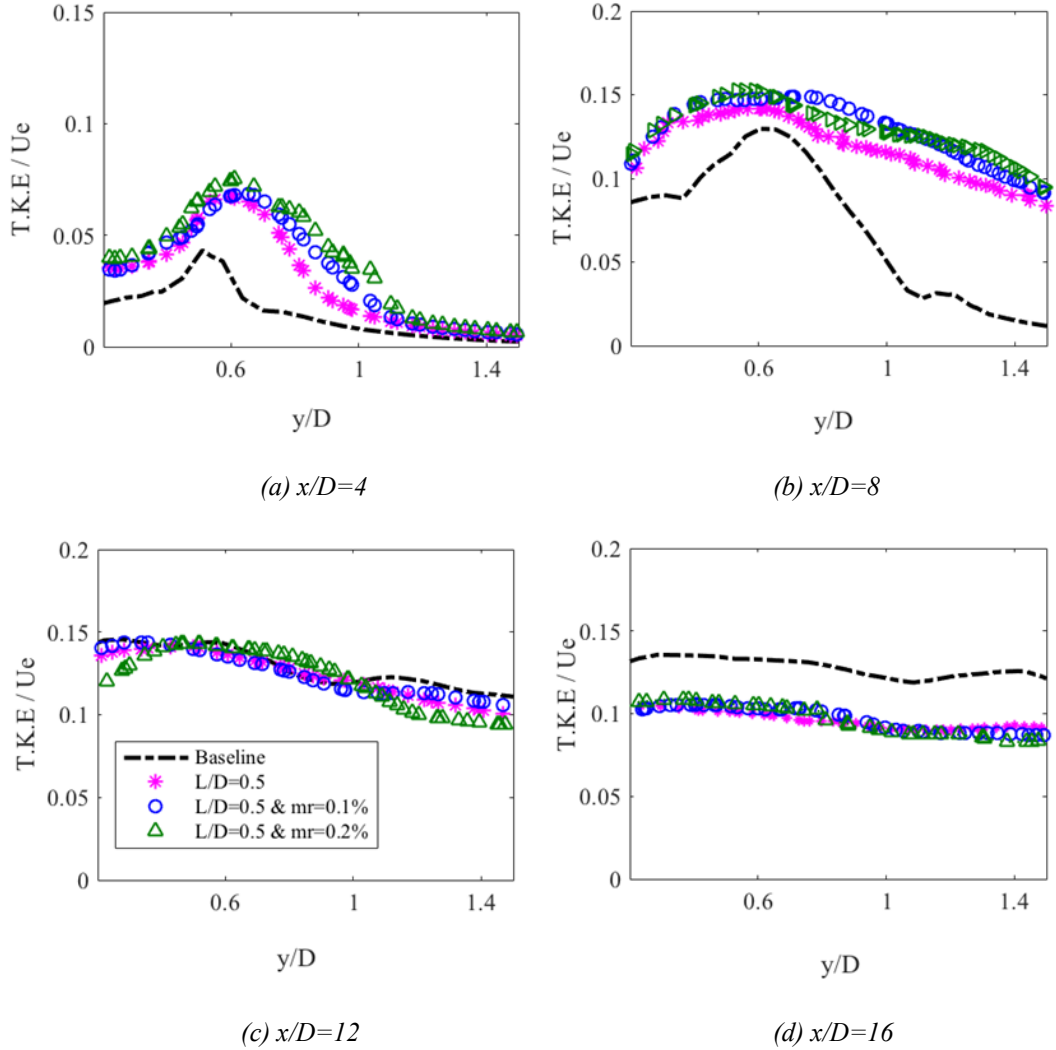

*(a) x/D=4*  *(b) x/D=8*

*(c) x/D=12*  *(d) x/D=16*

**Fig. 14: Profiles of Turbulent Kinetic Energy (T.K.E/$U_e$), comparing the baseline case to the cases with L/D=0.5**

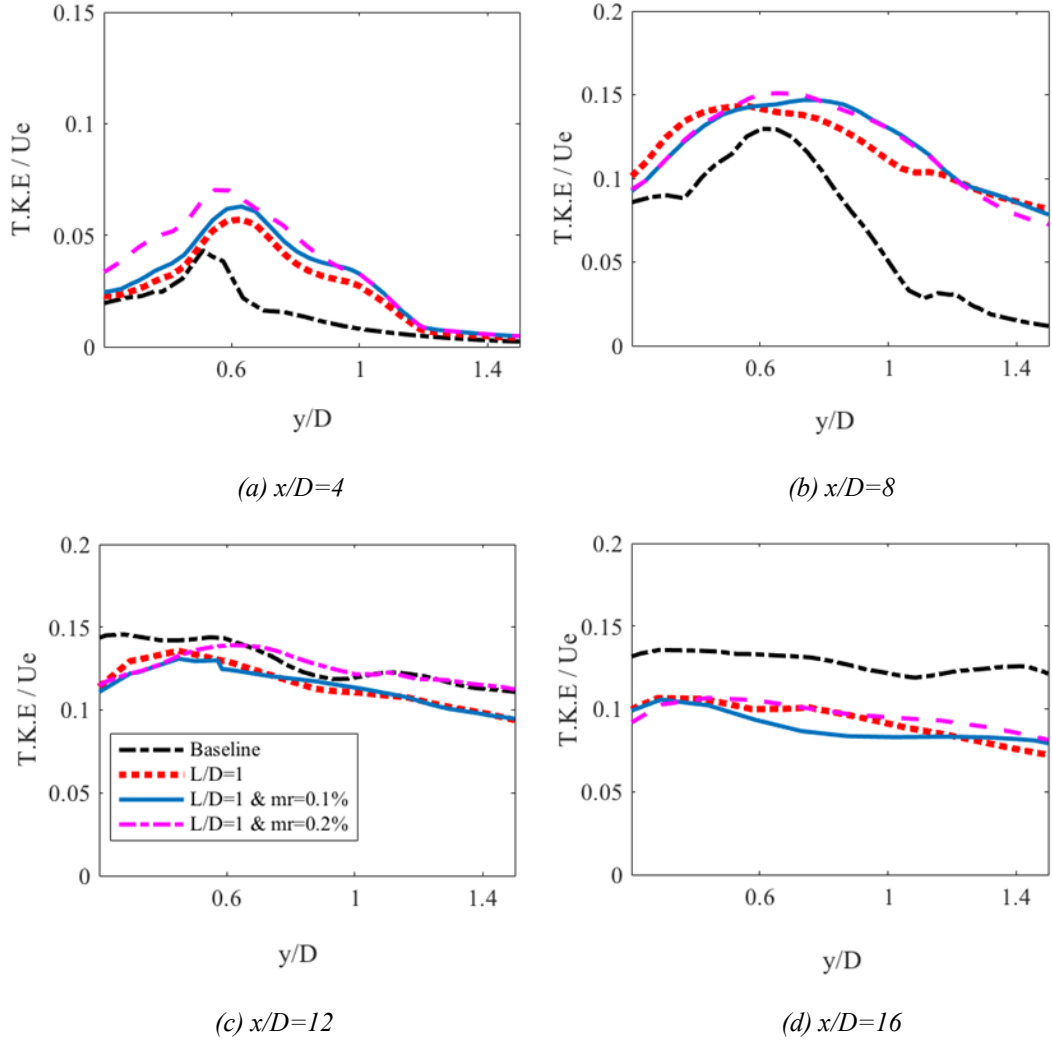

**Fig. 15: Profiles of Turbulent Kinetic Energy (T.K.E/$U_e$), comparing the baseline case to the cases with L/D=1**

### 4.3. Mean Flow Field Characteristics

#### 4.3.1. Jet Flow Development

The jet flow development in close proximity to the injection location is examined in Fig. 16. The contour plots of mean velocity for the cases involving fluidic injection with $m_r$ = 0.1% and 0.2% are compared to those for the baseline case. Due to the similarity of the velocity profiles for both microjet tube lengths, the results are shown for injection at $L/D$ = 1 and two jet momentum ratios. The flow field evolution downstream of the injection location shows the effects of the three-dimensional structures and the counter rotating vortex pairs on the mean velocity profiles for the cases with fluidic injection. This effect is more pronounced for $m_r = 0.2\%$, where higher injection penetration, as well as a growth in size of vortices are observed. Furthermore, a minor growth in the shear layer thickness at these distances for the cases with fluidic injection is visible. The effects of DMFI system on shear layer development are further discussed in the following sections.

Figures 17(a) and 17(b) compare the streamwise mean velocity profiles of the baseline case near the nozzle centerline at $y/D = 0.1$, to the cases with $L/D = 0.5$ and $L/D = 1$, respectively. Despite the minor influence of the presence of microjet tube on the mean flow condition at the nozzle exit, reduction of the shock cell strength as well as the length of the jet core is notable in downstream direction for those cases involving microjet tube compared to the baseline case. For the cases involving fluidic injection, the effect of microjet injection is immediately visible downstream of the injection location, as the alteration of the shock cell structures, in addition to the weakening of the shock cell strength, are observed. It should be noted that the effect of the boundary layer growth over the tube surface as well as the viscous wake behind the tube play important roles in shortening of the jet core length and enhancement of the jet mixing near the centerline [20]. All of these, in addition to the reduction of downstream turbulence production, which was discussed in the previous section, are expected to affect the turbulent mixing and the shock associated noise components of the jet.

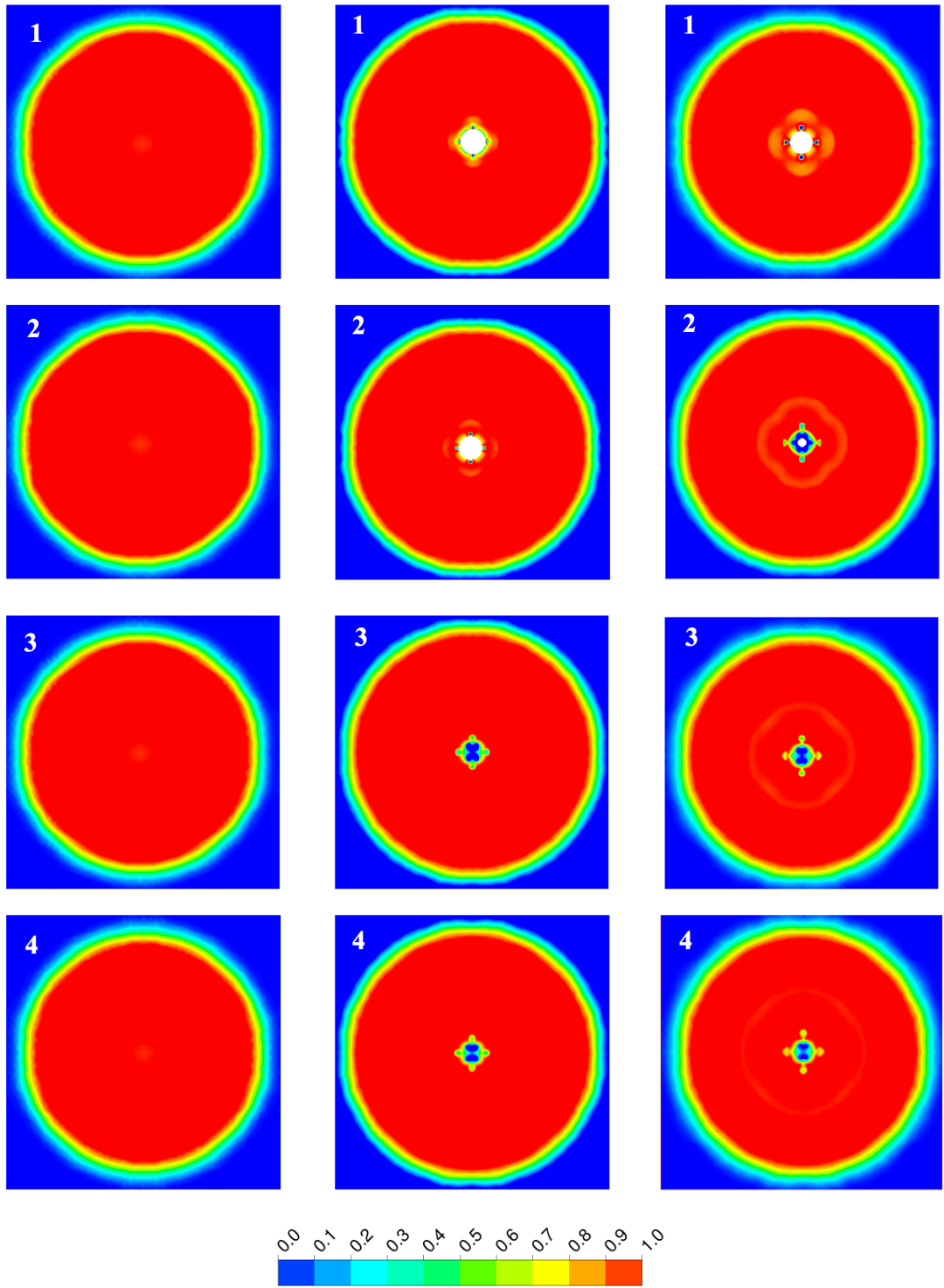

*(a) Baseline*  *(b) mr=0.1%*  *(c) mr=0.2%*

**Fig.16: Contour plots of mean velocity ($u_x/U_e$) for L/D=1, demonstrating the flow field evolution in y-z plane downstream of injection location. 1, 2, 3 and 4 correpond to *x/D*=1, *x/D*=1.05, *x/D*=1.1,and *x/D*=1.2, respectively.**

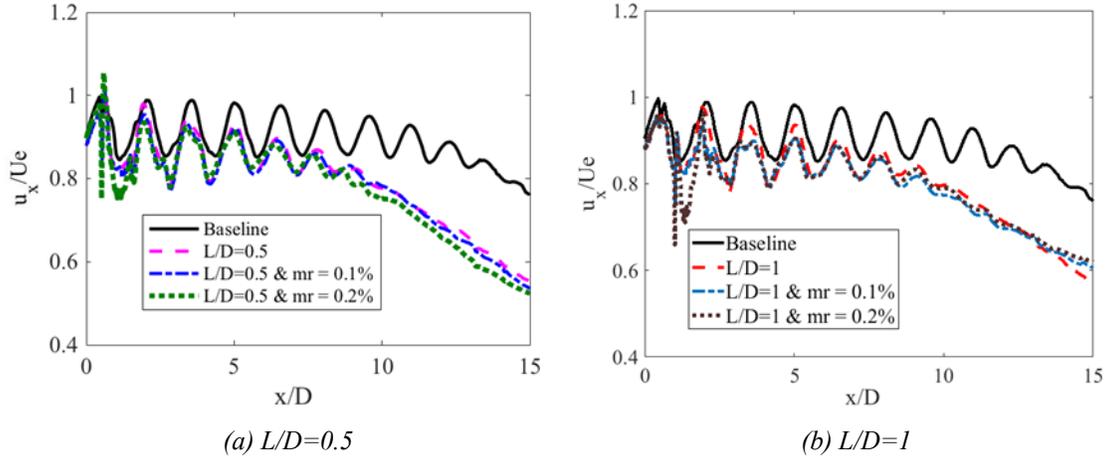

*(a) L/D=0.5*            *(b) L/D=1*

**Fig. 17: Comparison of streamwise mean velocity profiles ($u_x/U_e$) along y/D=0.1**

**4.3.2. Shear Layer Spreading**

The shear layer thickness, $\delta$, is examined through quantifying the upper and lower boundaries ($r_{0.1U_e}$ *and* $r_{0.9U_e}$) of the shear layer, where the mean velocities are approximately $0.1 U_e$ and $0.9\ U_e$ [9, 25], respectively, i.e.:

$$\delta = r_{0.1U_e} - r_{0.9U_e} \qquad (20)$$

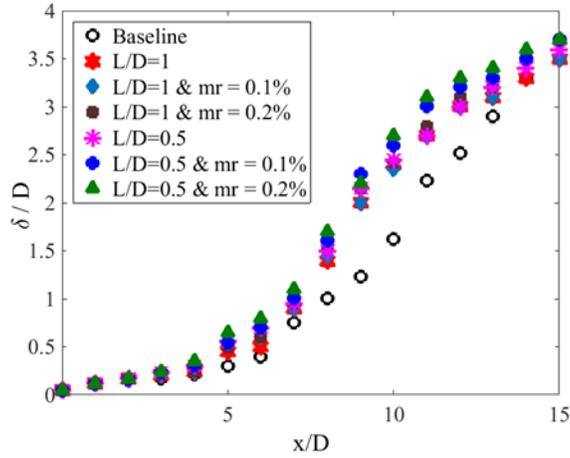

**Fig. 18: Comparing jet shear layer thickness of baseline case to the cases involving microjet tube**

As shown in Fig. 18, all cases involving microjet tube result in thickening of the shear layer compared to the baseline case. For all cases with microjet tube, thickness of the mixing layer retains higher values compared to the baseline case. Among the cases with the fluidic injection, those with injection at $L/D = 0.5$ demonstrate a higher growth of the shear layer thickness, which could be due to the enhanced early jet mixing at $x/D = 0.5$. To further investigate the growth rate of the jet shear layer, the jet half radius is considered in the streamwise direction. The jet half radius ($r_{0.5}$) is defined as the position, where the mean axial velocity is half of the centerline velocity [25]. Therefore, the growth rate of the shear layer can be estimated using the following expression:

$$\delta' = \frac{dr_{0.5}}{dx} \tag{21}$$

The spreading rate of the shear layer is examined through Eq. 21 for all cases and results are illustrated in Fig. 19 using a linear least-square fit, applied for two regions of the jet flow corresponding to $0 < x/D < 6$ and $6 < x/D < 15$. All cases involving the microjet tube increase the shear layer spreading rates compared to the baseline case, particularly for $x/D > 3$. The highest increase in the shear layer spread is, however, seen for those cases with $m_r = 0.2\%$, which approximately show 25% of growth. The spreading rate is further examined using the empirical relation developed by Papamoschou and Roshko [39]. According to this proposed relation, the growth rate of a compressible shear layer for a "single jet flow" with a quiescent surrounding ambient is obtained using the experimental curve-fitted data of Refs. [39-41] as

$$\delta' = \left[0.14\left(1+\sqrt{s}\right)\left(0.23+0.77e^{-3.5M_c^2}\right)\right] \tag{22}$$

with $M_c$ being the convective Mach number, and $s = \rho_\infty / \rho$. The above-mentioned relation is used to calculate the shear layer spreading rate for the baseline case of the present computational study, which will be later compared to the one obtained from the profile of the jet half radius (Eq. 21), as well as to the one which will be estimated using a theoretical planar shear layer model [42]. The planar shear layer model assumes that the shear layer grows with a constant rate, as shown in Fig. 20 and defined below:

$$\delta' \sim \frac{D}{L_c} \quad (23)$$

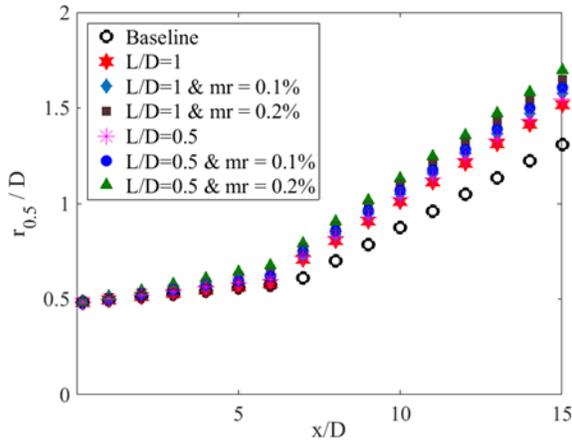

**Fig. 19: Comparing shear layer spreading rate of baseline case to the cases involving microjet tube**

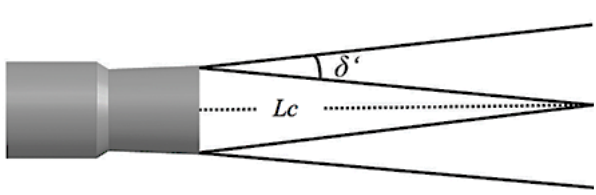

**Fig. 20: Jet core and shear layer development**

where $D$ represents the jet diameter at the exit of the nozzle, and $L_c$ is the length of the jet potential core. In this study, the jet core length ($L_c$) for different cases is estimated using the jet mean velocity profiles, which were previously presented in Fig. 17. In this manner, the jet core length is approximated by considering the distance from the nozzle exit up to the location, where the centerline jet velocity decays to nearly 90% of the velocity at the nozzle exit. The shear layer spreading rate for the baseline case is calculated using the aforementioned theoretical and empirical model (Eq. 22) and compared to the one estimated by the jet half radius (Eq. 21). Comparing the spreading rates presented in Table 1 for the baseline case shows that there is an overall agreement among the predicted results. As expected, the predicted value by the planar shear layer model (Eq. 23) is in a better agreement with the one estimated through the jet half radius, compared to the predicted growth rate by the empirical relation of Eq. 22. In a similar vein, the shear layer spreading rates for the cases involving microjet tube are predicted using Eq. 21, and the result are compared in Table 2. Examining the spreading rates indicates that all cases incorporating microjet are effective in increasing the rate of the shear layer growth compared to the baseline case.

| $\delta'$ | Eq. 21 | Eq. 22 | Eq. 23 |
|---|---|---|---|
| **Baseline** | 0.081 | 0.072 | 0.077 |

**Table 1: Comparison of shear layer growth rates for the baseline case**

| $\delta'$ | Baseline | L/D=1 | L/D=1 mr=0.1% | L/D=1 mr=0.2% | L/D=0.5 | L/D=0.5 mr=0.1% | L/D=0.5 mr=0.2% |
|---|---|---|---|---|---|---|---|
| **Eq. 21** | 0.081 | 0.101 | 0.106 | 0.111 | 0.103 | 0.107 | 0.113 |

**Table 2: Comparison of shear layer growth rates for the baseline case and cases involving microjet tube**

### 4.3.3. Flow Entrainment

As demonstrated in the previous sections, shear layer of the DMFI equipped jet has been modified considerably, with both arrangements showing increased spreading of the shear layer. As such, it is expected that the growth of the shear layer thickness influences the entrainment of the irrotational ambient flow into the shear layer of the jet. In order to examine the effect of DMFI system on the flow entrainment, the volume fluxes of the jet are calculated. The volume flux of the jet per unit thickness can be estimated for the jet potential core region as well as for the developed region as presented below:

$$q = \int_{-\infty}^{+\infty} \bar{u}(r,x) 2\pi r \, dr \tag{24}$$

Velocity profiles for the two regions of the jet are defined using Warren's analysis [43] for compressible turbulent jet mixing, based on the eddy viscosity formulation. It should be mentioned that the velocity profiles from the present numerical study are compared with those from Warren's analysis and due to the consistency of the results at most locations, and in order to simplify the analysis, the velocity profiles from Warren's analysis are used. The velocity profile for the jet core region is defined as:

$$\bar{u} = \bar{u}_1 \exp\left(-0.6932 \frac{r^2 - r_i^2}{r_{0.5}^2 - r_i^2}\right) \tag{25}$$

where $r_i$ refers to distance from the jet centerline to the edge of the potential core, and $\bar{u} = u / U_e$ is the jet velocity. In the developed region, where the jet potential core ends and the two ends of the shear layer merge together, the velocity profile is calculated as:

$$\bar{u} = \bar{u}_c \exp\left(-0.6932 \frac{r^2}{r_{0.5}^2}\right) \tag{26}$$

where, $\bar{u}_c = u_c / U_e$ is the jet centerline velocity in the developed region. Replacing the above-mentioned velocity profiles into Eq. 24 gives the flux of the jet for the jet core region as:

$$q = \frac{\bar{u}_1 \pi}{0.6932} (r_{0.5}^2 - r_i^2) \tag{27}$$

whereas for the developed region we have:

$$q = \frac{\bar{u}_c \pi}{0.6932} (r_{0.5}^2) \tag{28}$$

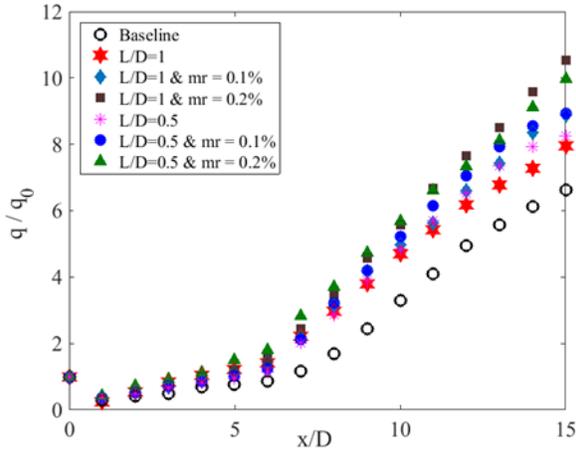

**Fig.21: Comparing jet volume flux ratio of baseline case to the cases involving microjet tube**

The volume fluxes of the jet are normalized by the volume flux at the nozzle exit, $q_0$, and comparisons are presented for 7 cases in Fig. 21. As seen in Fig. 21, the entrainment ratios near the nozzle exit assume very close values for all cases, but further downstream the entrainment ratios increase for the cases involving microjet tube compared to the baseline case. Among all cases examined including the cases with microjet tube and the baseline, those cases involving fluidic injection with $m_r = 0.2\%$ demonstrate higher flux ratios due to the higher shear layer spreading rates for these cases. Overall examination of the results indicates that DMFI system enhances the jet shear layer spreading, increases the flow entrainment from the surrounding ambient into the jet shear layer region, and hence, promotes the mixing in both near field and far field locations.

## 5. Conclusion

Flow field results of an underexpanded supersonic jet using a microjet tube with and without fluidic injection are investigated using a DDES methodology. The validations cases demonstrated that the DDES model implemented in the present study can adequately capture the mean and turbulent features of the flow field. Comparison of the results of the baseline case with the cases using DMFI showed that the DMFI system can be a viable method of jet mixing enhancement with a potential for noise reduction using minimal amounts of fluidic injection. Examination of the flow field features in presence of DMFI system indicate that increasing the jet momentum ratio results in stronger penetration of the fluidic injection into the supersonic jet accompanied by an increase in the strength and steepness of the injection induced bow shocks. Overall, arrangements with different microjet tube lengths and fluidic injection rates indicated effectiveness of DMFI on promoting the mixing characteristics of the shear layer with varying degrees of success. This finding, along with the fact that fluidic injection downstream of the nozzle exit is capable of targeting the regions of the shear layer with high susceptibility to alteration of the mixing characteristics, manifests the DMFI system as an effective method of mixing augmentation with the potential for suppressing the supersonic jet noise radiation. Studying jet noise reduction as well as the scaling of results to applications in larger jets are not within the scope of this paper and require follow-on studies.

## Acknowledgments

This research was supported by the Donald D. King Graduate Fellowship from NYU Tandon School of Engineering.